# On the Modification and Revocation of Open Source Licenses

Paul Gagnon, Partner, BCF
Misha Benjamin, Partner, BCF
Justine Gauthier, General Counsel & Head of AI Governance, Mila
Catherine Régis, Université de Montréal, Mila and IVADO,
Jenny Lee, Responsible AI Licensing
Alexei Nordell-Markovits[1,2]

Abstract

*Historically, open source commitments have been deemed irrevocable once materials are released under open source licenses. In this paper, the authors argue for the creation of a subset of rights that allows open source contributors to force users to (i) update to the most recent version of a model, (ii) accept new use case restrictions, or even (iii) cease using the software entirely. While this would be a departure from the traditional open source approach, the legal, reputational and moral risks related to open-sourcing AI models could justify contributors having more control over downstream uses. Recent legislative changes have also opened the door to liability of open source contributors in certain cases. The authors believe that contributors would welcome the ability to ensure that downstream users are implementing updates that address issues like bias, guardrail workarounds or adversarial attacks on their contributions. Finally, this paper addresses how this license category would interplay with RAIL licenses, and how it should be operationalized and adopted by key stakeholders such as OSS platforms and scanning tools.*

---

[1] The authors wish to thank Ninette AbouJamra, BCF, for her contributions.
[2] The authors can be reached for comments or questions at misha.benjamin@bcf.ca or paul.gagnon@bcf.ca.

1. Introduction

This Article deals with the modification and revocation of licenses by contributors, and puts forward a mechanism through which contributors can revoke licensed materials or add licensing restrictions in a manner consistent with the values of openness, scientific progress and accountability where there are demonstrable issues arising with the licensed materials.

Open source software (OSS) is a key driver of innovation and progress in the field of AI, and has been central to its evolution for decades. The recent releases of Large Language Models (LLMs) under OSS licenses has stoked the fire of AI innovation, creating technical and scientific advancements at a staggering pace. Public releases of LLMs and other materials has led to necessary scrutiny, probing and testing of materials made available, leading to well-documented cases of bias, privacy harms, security issues and other harms[3]. Some have even called for open source LLMs to be limited by regulation[4]. Commonly-used OSS licenses and their underlying conceptual frameworks fail to apprehend these issues, as they stem from "normal course" software development that does not consider AI's properties.

When materials are made available under OSS licenses, contributors cannot later retract permissions or rights they grant under any circumstance[5]. Even if materials are withdrawn from public platforms where they are hosted, contributors cannot revoke the rights of those who continue to make use of the materials, nor even stop licensees from re-sharing the materials or modifications made to them. In fact, Microsoft recently released its Wizard LM 2 model under an Apache license on Hugging Face before promptly deleting it as it had not undergone toxicity testing. Despite the deletion, the model had been downloaded and reshared, and Microsoft cannot retract that license[6].

However, the authors believe that there are clear cases where, in good faith and consistent with principles of non-discrimination, contributors should be entitled to later revoke or modify permissions to use materials. These include unforeseen and nefarious uses, emergent properties (including "sleeper" behaviour), output deficiencies, guardrail deficiencies, vulnerability to adversarial attacks that either disclose training data or allow data poisoning[7], model pipeline deficiencies and regulatory or legal changes. The suggested mechanism would enable this.

---

[3] A non-exhaustive list of these incidents is available at "Welcome to the Artificial Intelligence Database", AI Incident Database, 2024, https://incidentdatabase.ai/. From students being wrongfully suspended because test software based on open-source software could not recognize them, (See Mitchell Clark, "Students of color are getting flagged to their teachers because testing software can't see them", *The Verge*, April 20, 2021, ias-issues-opencv-facial-detection-schools-tests-remote-learning.) to unintended the inclusion of CSAM in open datasets (See David Thiel, "Investigation Finds AI Image Generation Models Trained on Child Abuse", *Stanford Cyber Policy Centre*, December 20, 2023, https://cyber.fsi.stanford.edu/news/investigation-finds-ai-image-generation-models-trained-child-abuse.) The examples are legion.

[4] David Evan Harris, "Open source AI Is Uniquely Dangerous" *IEEE Spectrum*, January 12, 2024, https://spectrum.ieee.org/open-source-ai-2666932122.

[5] "Open Source Software Licences User Guide", The Open University, accessed May 2024, https://www5.open.ac.uk/library-research-support/sites/www.open.ac.uk.library-research-support/files/files/Open%20Source%20Software%20Licences%20User%20Guide%20V4%20-%20clean%20copy(1).pdf.

[6] Kate Erwin, "Microsoft Takes AI Model Offline After Devs Fail to Conduct "Toxicity Testing"", *PCMag*, April 16, 2024, https://www.pcmag.com/news/microsoft-takes-ai-model-offline-after-devs-fail-to-conduct-toxicity-testing.

The mechanism put forward empowers contributors to: 1) impose use of a newer release or improved version of the materials (thus ceasing the use of previous versions); 2) add additional restrictions to the license terms; or 3) fully revoke the license and prohibit use of the materials altogether.

The license change mechanism put forward in this article addresses a conceptual gap in OSS licenses by allowing contributors to take into account downstream issues that appear once materials are released as well as emerging legal requirements such as the EU AI Act, Canada's Bill C-27 or any of the myriad AI acts that are being proposed in the US[8], many of which have or may in the future have provisions directly applicable to open-source content[9]. The framework put forward in this Article better helps assess individual liability and the liability of platforms hosting AI models and materials. Given emerging changes from a legal and regulatory standpoint, contributors stand to gain by obtaining a mechanism to manage and mitigate their own liability in the face of such changes. Platforms hosting AI models also benefit, in that the current environment creates uncertain obligations on their part to adopt and enforce guidelines against harmful AI models amidst an uncertain legal background.

This Article also puts forward draft language for comment and use in licenses for contributors. The goal is for contributors to make use of the mechanism described in this article by openly stating their intent to do so in the license(s) they select, and address restrictions on an ongoing basis. This language is meant to be compatible with common open source licenses, as well as emerging initiatives such as RAIL licenses[10].

---

[7] James X. Dempsey, " Generative AI: The Security and Privacy Risks of Large Language Models" *Netchoice*, April 2024, https://netchoice.org/wp-content/uploads/2023/04/Dempsey-AI-Paper_LLMs-Security-and-Privacy-Risks_April-2023.pdf.

[8] Goli Mahdavi, Amy de La Lama, Christian M. Auty, "US State-by-State Ai legislation Snapshot", *BCLP Client Intelligent*, February 12, 2024, https://www.bclplaw.com/en-US/events-insights-news/2023-state-by-state-artificial-intelligence-legislation-snapshot.html.

[9] See Section 5 for more detail. For instance, the EU AI Act has specific provisions (See Daniel Castro, "The EU's AI Act Creates Regulatory Complexity for Open-Source AI", *Center for Data Innovation,* March 04, 2024, https://datainnovation.org/2024/03/the-eus-ai-act-creates-regulatory-complexity-for-open-source-ai/#:~:text=While%20the%20final%20text%20of,source%20AI%20in%20the%20EU.). The NTIA in the US is exploring regulation of "Dual-Use foundation Models with Widely Available Model weights" which would apply to open source models, (See "Dual Use Foundation Artificial Intelligence Models with Widely Available Model Weights", National Telecommunications and Information Administration, Unites States department of Commerce, March 27, 2024, https://www.ntia.gov/federal-register-notice/2024/dual-use-foundation-artificial-intelligence-models-widely-available.) among others.

[10] "AI Licenses", Responsible AI Licenses, accessed May 2024, https://www.licenses.ai/ai-licenses.

Simply put, the authors believe that a clear framework to modify or revoke licenses to AI models and materials in clear and limited cases can further strengthen robust AI development and promote responsible use of AI.

2. New License Mechanism

The basket of rights proposed by the authors (the "Modifiable"-type licenses) contain three key elements. These three elements should be included as a block, and are not modular. This section puts forward the rationale and an example of the application of these elements. Proposed text for these elements is included in Annex A.

i. The right to modify the applicable license to include additional restrictions

The first right reserved in the Modifiable-type licenses is the right to change the license to include additional use restrictions, subject to the qualifying provisions below. Given that transparency is key to adoption by the open source community, it is necessary that this right includes the ability to only add clear use restrictions, rather than subjective or capricious additional conditions. For example, a condition to not use the licensed materials for "evil purposes" or "wrong"would be too subjectively determined to be valid. Capricious conditions imposing precise standards hampering use, such as prescribing a specific technological environment in which an AI model would be deployed and trained, would also not be deemed a valid restriction. Hence, additional restrictions must not change the nature of the given license itself and these restrictions should be clear and determinable by downstream users..

Additionally, contributors could wish to later add conditions to the license text intended to address certain issues arising after release of the materials. For example, a contributor could require that, prior to use, an algorithmic impact assessment be conducted to ensure due diligence in addressing the potential issues arising from use of the materials. Conditions would also need to meet the standards and qualifying provisions set out in the Article.The Qualifying provisions subsection discusses examples of non-permissible modifications.

ii. The right to revoke the license

The second right reserved in the Modifiable-type licenses is the right to withdraw the license, subject to the qualifying provisions set out below. This right serves a dual purpose: it allows the contributor to completely revoke the license so that the contribution can no longer be used, or to revoke the license and offer a new version of the contribution under the same license. In keeping with the principle of uniformity discussed above, the intent is that this revocation can only be used against all licensees, and not be used against specific licensees or categories of licensees only.

### iii. Qualifying provisions

Given the potential impact of modifying licenses, exercising this right should be constrained by the following qualifying provisions. These principles help ensure that the rights found in Modifiable-type licenses are exercised in good faith by contributors. For example, it would not be permissible to modify the license to create any redistribution obligations that did not already exist or to require a royalty, even if tied to certain use cases only. If a contributor wishes to require royalties or add additional license models, then a traditional fork and relicense would be the given mechanism. The Modifiable-type licenses could not be used to require a royalty for a previously free model where no additional value was added by the licensor, nor to revoke rights from a competitor or to simply remove the open-source license in order to start charging for a similar product.

The Modifiable-type licenses do not replace these dynamics, which all require the contributor to maintain the rights granted to licensees to the materials first licensed.

### a. Uniform and non-discriminatory application

The rights in the Modifiable-type licenses can only be exercised on a uniform basis across potential licensees and use cases, without discrimination or "singling out". If a modification is imposed, it is required that it be done against all similarly-situated licensees, and not against certain individuals or companies or classes of people. For instance, the language is not intended to create a withdrawal mechanism on a retaliatory basis against people who may be in litigation against the contributor, such as was the case with the React license when it was originally published[11]. For the mechanism in this Article to be consistent with open source principles, the modification itself as well as its enforcement need to be applied on a non-retaliatory and uniform basis.

As for restrictions based on territories or countries, it is self-evident that license changes meant to exclude or address specific geographies are subject to more subjective standards that may be inconsistent with the principles underlying open source. That being said, there are conceptually sound exclusions that could be consistent with the framework described in this Article. For example, a certain territory could be excluded in good faith based on the potential liability a contributor cannot exclude by law, or if legislation requires this contributor to assist or otherwise participate in audits or approval processes. In such cases, if documented and based in good faith, M-licenses could accommodate downstream changes impacting available geographies. By way of comparison, this could be similar to export control clauses, which in some cases have been deemed consistent with open source principles[12].

---

[11] Dennis Walsh, "React, Facebook, and the Revocable Patent License. Why it's a Paper", *Medium*, July 18, 2017, https://medium.com/@dwalsh.sdlr/react-facebook-and-the-revokable-patent-license-why-its-a-paper-25c40c50b562. Note however that a contributor could choose an open source license with commonly accepted retaliation mechanisms such as the Apache v2 or GPL v3 license, and subsequently add a modification under the principles proposed under this Article. The goal is for this modification not to be targeted or retaliatory.

### b. Good faith obligation

The rights in the Modifiable-type licenses can only be exercised in good faith, with a view towards avoiding potential societal harm (such as discrimination, human rights issues, severe quality issues, bodily harm) or liability caused by the continued use of the contribution. The language in Annex A specifically calls this requirement out. This means, for instance, that a use restriction cannot be imposed simply because the contributor wishes to start charging for use in a certain field, or because a competitor is using the contribution in that field. However, this good faith obligation should not be understood to mean that a contributor cannot later monetize a given field or use-case by adding a use restriction. This kind of modification would be implemented by a normal course fork of the license and materials. It is also essential to the nature of good faith exercise of these rights that they not be used in retaliation against licensees for reasons other than preventing future harm or liability.

A contributor can choose to document their assumptions and good faith basis to modify the license terms. For example, release notes can add context, or the contributor can point to other sources to document the basis for modification. The aim of Modifiable-type licenses is for contributors to modify when they hold reasonable, good faith bases to do so.

### c. Flow-Down of Revocation/Modification Notices

It is essential to the nature of the Modifiable-type licenses that once a contributor revokes or modifies rights related to a contribution, this change uniformly affects all downstream licensees. Therefore, anyone who distributes the contribution will also have the obligation to revoke and/or replace the license, and notify its licensees at least in the same way it distributes its contribution (e.g.. placing it in the Github files). The authors understand that based purely on the license files it would be unlikely that all downstream licenses would be revoked/modified. Section 7 contains a more detailed discussion regarding how this is likely to be operationalized.

### d. Remediation period

Contributors should implement and state a remediation period for licensees to enact the changes (in case of a modification) or to switch away from the materials (in case of a revocation). The remediation period should be proportional to the underlying issues addressed by the modification or revocation, as well as to the expected technical difficulty of implementing the modification.

The text proposed in Annex A has been drafted with a view towards compatibility with the most commonly used open source licenses. It is the intention of the authors that these open source licenses would each adopt a variant that includes a version of this text (ex. M-Apache 2.0 or M-MPL).

---

[12] "What is free software?", GNU operating System, *Free Software Foundation*, accessed May 2024, https://www.gnu.org/philosophy/po/free-sw.zh-tw-en.html.

### e. Application of Modifiable-license principles

By way of example, let us consider a contributor who decides to open source a model that analyzes body movement with the goal of allowing users to use it to track and prevent workplace accidents, including through a feature that analyzes gait in individuals. Upon release, the contributor chooses a given open source license, without any use-restrictions or other terms, and also states their intent to have this license be Modifiable.

Subsequent to releasing the model, the contributor then learns that the software is being used to conduct mass surveillance based on gait analysis. The contributor could modify the license and ban this use case across the board, without singling out any particular actors making use of the software in this way. If the contributor also learned that the gait analysis was being used to identify which workers need a break, but that the model significantly disadvantaged women, then the contributor could both (i) ban the use case of gait analysis because of possible harms of the model malfunctioning, and (ii) provide services or complimentary software that addresses the underlying questions of bias found in the model. Lastly, if the contributor believes that there are too many potential abuses of their model, or if the model was found to contain "sleeper"-type behaviour, revocation of the license could be considered. In all cases, the contributor would give licensees 30 days to implement the changes.

If undertaken under a Modifiable license (or M-License), these actions would be consistent with the principles stated in this Article:
- Uniform application - the contributor is not singling out any specific person or entity.
- Good faith - the contributor is making the modifications on a good faith basis, based on documented issues.
- Flow-down - the contributor expects downstream users to implement the changes to their own licensees.
- Remediation period - the contributor gives a reasonable delay to implement the changes.

### iv. Adoption and taxonomy

The authors suggest a simple and effective disclosure of the Modifiable-type licenses: adding an "M" before the short-form of the given license, or "Modifiable" when written in the long-form. The potential for modification of a given license bears consequence on licensees, so it is necessary that this be disclosed clearly and conspicuously. This taxonomy could be modified should there be consensus to do so subsequent to the publication of this Article.

Adoption of Modifiable-type licenses is arguably not unlike forks through which contributors elect a different license or opt for a dual-licensing model. Indeed, licensees ought to be mindful of licensing considerations for the open-source components they use, more so when such components are distributed by them to other downstream users. All users should avoid automatically pulling from third-party repositories without regularly confirming applicable licenses. As such, M-type licenses would be not different from other forks that result in changes in license terms.

Exercising due diligence and regularly monitoring builds to ensure license compliance is best practice, whether M-type licenses are used or not. The tools and processes to do so are available[13] and can be combined with cybersecurity considerations to scan and monitor builds. The principles underlying Modifiable-type licenses aim to foster transparency and good faith such that contributors can pull back contributions that represent valid risk. In turn, licensees are duly informed and can implement changes responsibly, and pass these along to their own downstream users. Without the framework of Modifiable-type licenses, the AI community implicitly accepts that "the genie cannot be put back in the bottle" - a defeatist approach that fails to take into account the well-documented risks discussed below.

As will be discussed below in Section 5, the legal landscape in the field of AI is evolving towards heightened responsibility for actors across the AI value chain. The authors hope that Modifiable-type licenses will become part of a culture of responsible AI, helping both contributors and licensees alike in striving to provide best in class AI models that aim for compliance in a fast-changing legal environment.

3.     Practical Implications

      i. Potential Chilling Effects

The authors accept that certain OSS users may avoid selecting or investing in open source software licensed under Modifiable-type licenses due to long term uncertainty of access and the work it could require to rip out and replace a code or a model if required. However, a consensus seems to be emerging that enterprise applications should be able to run on top of different LLMs[14] in order (i) to capture the latest innovations given the rapid pace of change in the industry[15], (ii) to utilize the model best adapted for a narrow task, and (iii) to protect against overreliance on one provider[16]. This does not mean that monitoring the latest developments and replacing underlying OSS as required will be trivial, but this trend towards LLM agnosticism coupled with standard OSS monitoring tools makes Modifiable-type licenses practical to build upon.

---

[13] A number of tools are already available to help monitor both license compliance and flag where an open-source module should be updated to the latest version because of known bugs or security issues, such as Blackduck. See "Black Duck Software Composition Analysis", Synopsys, accessed May 2024, https://www.synopsys.com/software-integrity/software-composition-analysis-tools/black-duck-sca.html. Also see "Secure Your Products from Repo to Release", Fossa, accessed May 2024, https://fossa.com/. Also see "Code Insight", FlexNet, accessed May 2024, https://www.revenera.com/software-composition-analysis/products/flexnet-code-insight.; among others.
[14] Lucy Mazalon, "AI Wars: How Salesforce's Agnostic LLM Approach Works", *SF Ben*, August 04, 2023, https://www.salesforceben.com/ai-wars-how-salesforces-agnostic-llm-approach-works/. Also see: Adrian Bridgwater, "LLM series - BlueFlame AI: Why we need to believe in LLM-agnosticism", *TechTarget*, December 2023, https://www.computerweekly.com/blog/CW-Developer-Network/LLM-series-BlueFlame-AI-Why-we-need-to-believe-in-LLM-agnosticism.
[15] Carl Franzen, "Consulting giant McKinsey unveils its own generative AI tool for employees: Lilli", *VentureBeat*, August 16, 2023, https://venturebeat.com/ai/consulting-giant-mckinsey-unveils-its-own-generative-ai-tool-for-employees-lilli.

On the contrary, there is an argument to be made that M-type licenses could encourage future contributions that would otherwise not be made given that, absent M-type licenses, contributors are bound by irrevocable commitments, notwithstanding potential liability or downstream issues.

### ii. Enterprise Policies

It is the expectation of the authors that where a modifiable-type contribution is made by a company, the company itself (and not the individual pushing the contribution) should have the sole ability to use those rights. This would imply certain best practices that would have to be either implemented or insisted upon, such as:
   a. Ensuring that all contributions are made from enterprise accounts, as opposed to individual accounts. This is generally the case anyway, but some smaller companies are known to be lax on these policies, especially when it comes to contributing to existing OSS communities.
   b. Having clear internal policies regarding who can make a decision to use a Modifiable-type right and permission setting that reflects those policies
   c. Creating internal committees or guilds that would advise on and/or decide when to trigger the Modifiable-type rights[17]
   d. Where companies want to create an active community around their OSS contributions, having public-facing policies on how and when it would trigger Modifiable-type rights, and even consider having non-company community members be part of the decision-making process

### 4. Unforeseen effects of AI contributions

The legal, reputational and moral risks related to the misuse of AI models, and generative AI models in particular, do not need to be restated. There are constant new reports[18] of harm caused by misuse of AI. From the use of open source models to create deep fake pornography[19], to rampant bias in the output of generative AI models[20], it is difficult for those making AI models available to the public to predict all of the ways in which it may be misused. A few categories include:

    a. Unforeseen uses: Despite due diligence prior to model release, there may be additional use cases and issues arising from use subsequent to release of open source systems. Although some licenses prohibit adverse uses generally such as do no harm licenses[21] or prohibit specific uses such as RAIL licenses[22], it is the authors' opinion that these licenses do not always properly allow contributors to address all possible prohibited uses. This is because prohibited uses are determined upon release, based on sometimes ambiguous definitions. Moreover, as the social acceptability of certain AI use cases evolves over time, it is virtually impossible for a contributor to know what they intend to allow and disallow in advance.

    b. Output deficiencies: Even where open source models are tested for output quality, including potential biases, this testing tends to focus on a narrow range of uses, be static and be based on certain key benchmarks that are known to have deficiencies[23]. This risk becomes greater the more the actual use deviates from the primary intended use for which the model was trained and tested.

    c. Guardrail deficiencies: Open source generative AI models generally have built in guardrails designed to prevent the model from outputting content that would be illegal (such as instruction on how to build explosive devices or synthesize drugs), immoral (such as create bigoted speech, defamation or misinformation) or otherwise contrary to the wishes of the contributor. However, we have seen time and again that these guardrails are either missing or can be easily bypassed using prompt engineering[24] or

---

[16] For instance, Walmart switched to an LLM-agnostic approach in the wake of the OpenAI leadership shakeup.
[17] In the authors' experience, the composition of these types of guilds is best when flexible and driven less by type of expertise than interest for the topic. Of course the guilds would have to bring in specific expertise (Ex. legal) as needed, but we these are first and foremost be led by members who are active within and care about the open source community.
[18] "Welcome to the Artificial Intelligence Database".

---

[19] Tatum Hunter, "AI porn is easy to make now. For women, that's a nightmare", *The Washington Post*, February 13, 2023, https://www.washingtonpost.com/technology/2023/02/13/ai-porn-deepfakes-women-consent/.

[20] Alexandra Sasha Luccioni, Christopher Akiki, Margaret Mitchell, Yacine Jernite, "Stable Bias: Analyzing Societal Representation in Diffusion Models" *Cornell University*, November 09, 2023, https://arxiv.org/abs/2303.11408.

[21] "Hippocratic License 3.0 (HL3): An Ethical License for open Source Communities", The Hippocratic License, accessed May 2024, https://firstdonoharm.dev/.

[22] Carlos Munoz Ferrandis, Danish Contractor, "The BigScience RAIL License", *BigScience*, accessed May 2024, https://bigscience.huggingface.co/blog/the-bigscience-rail-license.

[23] Mihir Parmar, Swaroop Mishra, Mor Geva, Chitta Baral, "Don't Blame the Annotator: Bias Already Starts in the Annotation Instruction", March 20, 2024, https://arxiv.org/pdf/2205.00415.pdf.

[24] Andy Zou, et al., "Universal and Transferable Adversarial Attacks on Aligned Language Models", Carnegie Mellon University, Center for AI Safety, Google DeepMind, Bosh Center for AI, December 20, 2023, https://arxiv.org/pdf/2307.15043.pdf. Also see: Frank Landymore, "Car dealership disturbed when its AI is caught offering Chevys for $1 each", The Byte, December 21, 2024, https://futurism.com/the-byte/car-dealership-ai.

training[25]. Moreover, adversarial attacks can result in models regurgitating training data[26], which may be in violation of applicable licenses or privacy laws. In fact, it has even been shown that employees can build in code that contains malicious "sleeper" behaviour that would not be discoverable during safety testing and red-teaming, and that would only emerge after distribution[27].

d. Model pipeline and data deficiencies: Increasingly, there is a focus on the entire model development chain, including how data is sourced and annotated. Unbeknownst to the contributor at the time of the contribution, their model may have been developed using slave or child labour[28], improperly gathered personal information or biased information. The model could also have started from a pre-trained model that was improperly licensed or trained on copyrighted material. The model could also have evidence of malicious "sleeper" behaviour or other technical issues compromising its integrity.

The authors believe that contributors would welcome a mechanism that allows them to either force users to update to the latest version of the software, accept new license restrictions or cease using the models altogether. This mechanism would empower contributors to engage with critical issues that can emerge, including the examples above.

5. Evolving legal environment

Adding to the risks related to the above unknowns is the rapidly evolving legal environment applicable to those who create and use AI. The goal of this paper is not to create a comprehensive review of applicable legislation, but the authors wish to highlight a few examples of current developments that illustrate why contributors may have good reasons to retain the ability to update terms or models, or withdraw them entirely. Without the mechanism described in this Article, contributors would potentially expose themselves to liability in certain cases. Minimally, this could discourage contributions to the dynamic culture of open source underlying advancements in AI, thus chilling the rapid pace of innovation.

a. EU AI Act: The EU AI Act is one of the most advanced comprehensive AI legislations and contains hefty fines for violations. In its original draft, it did not address open-source software or foundational models specifically, and it was unknown whether or not contributors of open source software would be required to comply with the Act based on end uses that were possible or that others made of contributions. As adopted, the EU AI Act addresses harms arising from open source AI, and does not provide a blanket exemption for open source AI. Indeed, if general-purpose models are released under an open source license, it may still be subject to the EU AI Act if it presents characteristics that qualify it under prohibited AI practices or are otherwise considered general-purpose models. The EU AI Act's scope presumably indirectly includes contributors making available open source software later incorporated by others in the EU. Lastly, the authors note that the language of the AI Act introduces uncertainty around "open source" as commonly understood as "AI components that are provided against a price or otherwise monetised, (...), should not benefit from the exceptions provided to free and open source AI components"[29].

---

[25] Tiernan Ray, "Generative AI can easily be made malicious despite guardrails, say scholars", *ZDNET*, November 29, 2023, https://www.zdnet.com/article/generative-ai-can-easily-be-made-malicious-despite-guardrails-say-scholars/.
[26] Eduard Kovacs, "Simple Attack Allowed Extraction of ChatGPT Training Data", *SecurityWeek*, December 1, 2023, https://www.securityweek.com/simple-attack-allowed-extraction-of-chatgpt-training-data/.
[27] Evan Hubinger et al., "Sleeper Agents: Training Deceptive LLMs That persist Through Safety Training", *Anthropic, Redwood Research, Mila Quebec Institute, University of Oxford, Alignment Research Center, Open Philanthropy, Apart Research,* January 17, 2024, https://arxiv.org/pdf/2401.05566.pdf.
[28] Niamh Rowe, "Underage Workers Are Training AI", *Wired*, November 15, 2023, https://wired.com/story/artificial-intelligence-data-labeling-children/.

Alongside the Act, the EU is also developing liability guidelines that may create avenues for civil liability. Although there is a high-level carve out of for open source software, there are exceptions to this carve out[30], and it is unclear how these exceptions may evolve both in this directive and similar ones in different jurisdictions.

b. EU General Product Safety Regulation (GPSR): this regulation overhauled the EU framework for consumer-facing products and services and also creates traceability and reporting requirements. It applies to products made available in the EU, including for free, and certain obligations extend to component-level and software embedded in products. This Regulation requires that firms commercializing products, including those embedding open source software, must conduct and document risk-analysis as well as corrective measures taken, with risk and danger being defined in terms of health and safety of consumers. The EU AI Act explicitly lists it as a backstop against AI released in the EU market that is not otherwise captured by the EU AI Act[31].

c. US Executive Orders: while various jurisdictions in the US are tabling legislation applicable to AI[32], there is also a patchwork of executive orders that may apply in surprising ways, and that are subject to change based on the administration in power. For example, the Executive Order on the Safe, Secure, and Trustworthy Development and Use of Artificial Intelligence[33] contains obligations that could apply to OSS contributors in a way they would not have predicted.

---

[29] Recital 103, EU GDPR, May 2018, https://www.privacy-regulation.eu/en/recital-103-GDPR.htm.
[30] Philipp Behrendt, Katie Chandler, "OSS liability- in light of the new Product Liability Directive" *TaylorWessing*, October 3, 2023, https://www.taylorwessing.com/en/interface/2023/open-source-software/oss-and-liability-in-light-of-the-new-product-liability-directive.
[31] Recital 166, EU GDPR, May 2018, https://www.privacy-regulation.eu/en/recital-166-GDPR.htm#:~:text=(166).
[32] IAPP Research and Insights, "Global AI Law and Policy tracker", *iapp ai governance center*, January 2024, https://iapp.org/media/pdf/resource_center/global_ai_legislation_tracker.pdf.
[33] Joseph R. Biden JR, "Executive Order on the Safe, Secure, and Trustworthy Development and Use of Artificial Intelligence", *The White House*, October 30, 2023, https://www.whitehouse.gov/briefing-room/presidential-actions/2023/10/30/executive-order-on-the-safe-secure-and-trustworthy-development-and-use-of-artificial-intelligence/.

d. Evolving enforcement landscape: Within the existing legal framework, the US FTC has shown a willingness to use its existing enforcement tools in ways that can have heavy impacts on AI developers. For instance, it has indicated[34] that intends to force companies to delete models that were trained on improper data, and retain them (otherwise known as algorithmic disgorgement). It is not currently know how a company's open sourcing of a model it was ordered to disgorge would affect that company's treatment by the FTC.

e. Ongoing case law: at the time of writing of this paper, there are a number of cases alleging that the use of copyrighted and/or personal information to train a model constitutes infringement of the IP rights inherent in that training data[35]. Although this question may eventually be solved by the US Supreme Court and/or by specific legislation, the authors believe that there will not be consensus on this issue for years to come due to the high stakes at play, the fact that different countries have different rules on fair use and that these models will be an integral part of the world economy by the time the cases reach the higher courts, especially at a global scale. It is unclear how the release of a model under a non-retractable open source license would affect the damages question should the complainants of these cases ultimately prevail, but the authors expect that the ability to withdraw the licenses would factor into damages arguments.

6. Modifiable-type licenses and open source

i. Irrevocability of OSS licenses

The revocation mechanism is required because existing OSS licenses do not confer the right for licensors to retract their contributions and restrict users from using them. Indeed, OSS licenses are grounded in core principles that ensure the freedom for licensees to use the materials without restriction, including from subsequent changes of terms by a licensor. Basically, if a licensee accesses materials under an OSS license, it can do so in perpetuity, under the same license terms. If subsequent versions are made available under a different license, this does not prevent a user from using the previous version under the previous license terms.

---

[34] Tonya Riley, "The FTC's biggest AI enforcement tool? Forcing companies to delete their algorithms", *Cyberscoop*, July 5, 2023, https://cyberscoop.com/ftc-algorithm-disgorgement-ai-regulation/.

[35] These include cases such as Getty Images v. Stability AI, a class action against Stability AI, Midjourney and others (See Andersen et al v. Stability AI Ltd. et al, https://imagegeneratorlitigation.com/pdf/00201/1-1-stable-diffusion-complaint.pdf.). Also see Michael M. Grynbaum, Ryan Mac, "New York Times Sues OpenAI and Microsoft over A.I. Use of Copyrighted Work", *The New York Times*, December 27, 2023, https://www.nytimes.com/2023/12/27/business/media/new-york-times-open-ai-microsoft-lawsuit.html. For an evolving list, also see Tiana Loving, "Current AI Copyright Cases - Part 1", *copyright alliance*, March 30, 2023, https://copyrightalliance.org/current-ai-copyright-cases-part-1/.

The Open Source Initiative puts forward ten criteria[36] to determine whether a license is deemed to be open source. The Free Software Foundation's approach is simpler and couches its definition in the existence of four freedoms, namely: (i) the freedom to run and copy the software; (ii) freedom to study, change and improve the software; (iii) freedom to distribute copies; and (iv) freedom to distribute a modified version. Open source is contrasted with proprietary software, where the licensor's ability to determine and change license terms is deemed to run counter to the ethos of freedom.

The Free Software Foundation recognizes that, in order for the freedoms underlying open source software "to be real, they must be permanent and irrevocable as long as you do nothing wrong", adding that "if the developer of the software has the power to revoke the license, or retroactively add restrictions to its terms, without your doing anything wrong to give cause, the software is not free"[37]. Couching revocability under moral grounds (i.e. doing something wrong) is not a straightforward approach, and so, from a legal standpoint, one must assume that what is considered "wrong" would be to breach the terms and conditions of a license. As for the Open Source Initiative, it does not state "irrevocability" as a core pillar of its "Open Source Definition"[38].

OSS licenses do not even systematically address revocability or license termination in case of a licensee's breach of its terms. Indeed, the grant of licenses under most OSS licenses is explicitly stated as being irrevocable. That being said, because OSS licenses constitute enforceable contracts, it is necessary that they be revocable in case of non-compliance with the given terms and conditions, however minimal or extensive those may be.

As for the language of commonly used OSS licenses, simpler licenses like the MIT or BSD licenses do not expressly address revocability at all. More exhaustive licenses like the Apache or GPL families of licenses address revocability, but only within certain circumstances that deal with a licensee's compliance with the license. For example, the Apache v2 license states that the underlying patent license is terminated in the event that the licensee institutes patent litigation against any third party which alleges that the licensed materials are infringing the licensee's patent(s). The GPL v3 license explicitly states that non-compliance of the license's terms results in automatic termination of the licensed rights, subject to a cure period.

Thus, commonly accepted principles in open source licensing only deal with licensee breaches and fail to address revocation of a license by the contributor. This important gap presents specific and clear risks with respect to AI.

---

[36] These are: 1) free redistribution; 2) access to source code and compiled code; 3) right to modify and create derived works; 4) limited ability for the licensor to restrict distribution of modified source code; 5) no discrimination against persons or groups; 6) no discrimination against fields of endeavor; 7) absence of additional license terms or requirements; 8) license cannot be specific to a product; 9) license cannot restrict use alongside other software; 10) license must be technology-neutral.
[37] "What is Free Software?".
[38] "The Open Source Definition", Open Source Initiative, February 16, 2024, https://opensource.org/osd/.

### ii. Liability under OSS licenses

a. Are disclaimers in OSS licenses enforceable?

There has been a consistent intent to shield contributors making open source commitments from liability arising from use of the open sourced materials. This is apparent from the language of most OSS licenses - the contributor's liability is explicitly disclaimed. This includes warranty disclaimers through which quality or fitness for purpose are expressly adopted. Users of OSS are familiar with these terms: there is a strong sense that "what you see is what you get" and that use of OSS is at the user's own volition and responsibility.

Adopting licenses that shield contributors from liability is coherent, and spurs innovation in that it is acceptable for individuals and organizations to expressly disclaim warranties and liabilities. After all, OSS is free to use, and if licensees are not satisfied, it remains entirely within their power to move away from OSS materials. If OSS contributors were held to strict quality standards or service levels to update and maintain their contributors, surely this would disincentivize contributions being made in the first place. There are of course active contributors and communities maintaining and addressing bugs and issues, as well as bringing forward improvements to the materials. But, these are not performed under any legal obligation as there are no legally binding commitments to contribute or improve.

That being said, it is far from certain that these disclaimers are legally enforceable. In many jurisdictions, it is not permissible to exclude one's liability by way of disclaimer. For example, in many common law jurisdictions, including a number of US states, disclaimers from gross negligence or fraud are not enforceable. In Quebec, no waivers against bodily harm can be enforced, and generally speaking, liability for intentional or gross fault (as defined under Quebec civil law).

Are warranty disclaimers and exclusions of liability unnecessary? Far from the authors to call into question such an important pillar of OSS contributions - it remains fundamental that contributors deliberately and openly state their wish to shield themselves from liability. But, knowing that this wish has its limits from a legal standpoint, there are no currently accepted means to revoke permissions to OSS contributions. So, even if contributors later wish to mitigate their liability, absent the mechanism described in this article, there are no commonly accepted means to do so.

This issue is compounded by the fact that the right to redistribution by downstream users is a key principle underpinning OSS licenses. Subject to not imposing additional restrictions onto other downstream users, the licensed materials can be re-shared, incorporated into a broader whole or on a stand-alone basis, "as is" and under the same license terms as those from which the licensee benefits. This exacerbates the initial contributors' potential liability, in a manner beyond their reach and control. It also creates additional liability for downstream users: if a known or expected issue arises with the licensed materials, their own liability may be triggered by their redistribution of the licensed materials. This is even more so where they have knowledge of the issues generating potential liability prior to redistribution.

Take, for example, a known cybersecurity issue with an open source library. If a licensee uses this library as part of another product or offering and fails to remedy a known breach, including by neglecting to use an updated version in which the underlying issue is fixed, the licensee could reasonably be considered imprudent, if not outright neglectful, if the licensee has clear and unequivocal knowledge of the issue and continues to redistribute.

The technical issues with AI models described above can be just as nefarious as cybersecurity issues. The modification and revocation mechanisms described in this Article constitute a tangible way for contributors to manage their liability, beyond the wishful thinking embodied in warranty and liability disclaimers.

b. Limited Application of Disclaimers

Where the standard disclaimers are enforceable, it should however be noted that they only apply to licensees of the licensed materials. This means that it may not be enforceable on users of a SaaS product that are not "using" the materials, whether or not they receive the OSS notice file. Although a separate disclaimer may apply to users of such a service, it is likely that such a disclaimer would not cover contributors of the OSS components underlying the SaaS product. Although it could be argued that there is a lack of privity between an OSS contributor and the end-user of a SaaS-based system that uses that contribution, emerging legislative guidance seems to point towards the creation of additional liability for different actors along the AI value chain, which do not as yet specifically exclude OSS contributions[39].

In addition, these disclaimers are not helpful to the extent that the issue with the contribution relates to the infringing contribution by a third party, such as improper use of training data or use of a model that was not properly licensed. Given that the disclaimers protect from suits against licensees only and do not provide for indemnification by such licensees, these disclaimers do not serve to protect a contributor from responsibility for the additional damages for every use of the contribution if the contribution is found to infringe third party rights.

a. RAIL licenses

Responsible AI Licenses (or "RAILs") were intended to provide a mechanism for AI developers to share AI "artifacts" (models, code, or data) subject to certain behavioral use restrictions. Prior to sharing a model with the community, for example, a developer could select a RAIL with specific restrictions that were important to such developers, in light of the artifact being licensed. For example, if the model could be used to identify bacterial diseases in humans, the RAIL restriction might include a term stating that the model should not be

---

[39] Although the EU product liability guidelines do contain exemptions for OSS, there are carve outs to those exemptions, such as for commercial activity, or where personal data is used in certain ways. For a list of relevant provisions see Daniel Castro, "The EU's AI Act Creates Regulatory Complexity for Open-Source AI".

***exclusively*** relied upon to diagnose an individual. The RAIL restriction would thus urge caution and double-checking by future users of the AI model.

The M-licenses discussed herein have a similar philosophical foundation as RAILs, in that both types of licenses wish to curb certain "bad" behaviors that are foreseeable by the licensors. The M-licenses perhaps have the advantage of encouraging ongoing attention by the licensors – that is, developers who choose to share code under OSS terms may opt to include M-licenses in order to later impose restrictions, based on circumstances as they unfold. The same mechanism may be used with RAILs, but only via amendment of the original RAIL terms.

7. Technical considerations for license changes

The application of M-licenses requires due consideration to technical and practical considerations, which mostly revolve around ensuring adequate flow of information to downstream users pertaining to a license change or revocation.

The git system of code versioning and development, used by most code sharing platforms like Github, Gitlab and Hugging Face, does not offer an effective way to "push" license modifications to downstream libraries and users making use of the M-licensed code. Additionally, git-based systems are decentralized and provide any user a "cloned" code repository from which to build additional systems, modify the licensed code and subsequently host or use the given code. There is no "call home" or "phone in" communication back to the initial M-licensed repository. Thus, practically speaking, there is a gap in automated enforcement of license changes and modifications. There is no effective "kill switch" in git-based systems. As such, at present, the application of M-licenses would focus on education of users and explicit "human readable" materials communicated alongside the license.

Additionally, even within a common platform (such as Github) there may be any number of forks or intermediate libraries between the original repository and the other downstream users. M-license changes or revocation should not be perceived as a "one to one" communication but as a "one to many", layered communication. Further compounding issues of communication is the fact that intermediate libraries may also have been abandoned by their maintainers, creating additional burdens for users to trace back source repositories that may be M-licensed and have been subject to changes.

A potential solution is for the license to require users to register to a lightweight information diffusion system (such as a newsletter). Such a system could preserve developer anonymity, would be relatively lightweight for both the original and downstream developers to build and use, and would allow easy downstream flow related to any license changes. The original library developer could potentially tie in any license changes to an automatic email message via their chosen release mechanism. That being said, such registration could be perceived as "gatekeeping" contrary to the ethos of open source software, in addition to raising compliance burdens for licensors, who then become responsible for managing email addresses and other potentially personal information protected by law. Ultimately, it is the responsibility of licensees to ensure compliance, and therefore contributors of M-licenses have no obligation to put in place the above mechanisms, but those who are particularly concerned can implement these mechanisms to ensure better compliance.  Platforms themselves could build notification mechanisms that could be automated and even AI-based. Finally, users could leverage existing code scanning tools that address license compliance as well as security concerns to effectively track and monitor license changes .

In the case of a modification or revocation to M-licensed code that is effectively communicated to a downstream user, the task of replacing or modifying licensed code is not always straightforward, specifically in the field of AI. Deployment of AI systems requires substantial testing and fine-tuning to meet performance requirements. Where code is replaced altogether due to a license change, the downstream user will want to ensure adequate if not equivalent or better performance of the now-replaced AI model. "Normal" changes of libraries in software remain impactful, but not altogether unwieldy given that the impacts of a change are measurable when the code compiles (or does not). For AI models and components, switching away carries high impacts, including with respect to testing and measuring performance and safety. The impact of requiring a change or license revocation itself cannot be underestimated. Thus, the time required to implement license changes should take into account the complexity of the required changes and be adjusted accordingly.  That said, potential liability for contributors and the harms described in this Article need to be balanced with the practical considerations of replacing or modifying code. The fact that license revocation and modification is impactful should not result in defeatism and laissez-faire approaches that ultimately create liability for contributors.

8. M-Licenses and platform obligations

While OSS licenses can be used in multiple contexts, it is through platforms where materials are openly distributed that OSS licenses live and become most apparent. The importance of platforms like GitHub, GitLab and HuggingFace cannot be understated in the world of AI. These platforms ensure broad and largely unfettered access to open source materials. They provide this service as would-be neutral third parties facilitating exchanges between users making materials available and those availing themselves of their rights under the given OSS licenses. Typically, the terms of service of third party platforms expressly disclaim their liability for materials they host and intend to shift liability for compliance to contributors and end users.

That said, platforms are not entirely immune to legal or ethical liability related to materials they host. Indeed, they could be compelled by competent courts to remove materials that are deemed illegal. This includes material infringing third-party intellectual property rights, privacy rights or other recognized rights or interests.  For example, if a database hosted on a platform contains personally identifiable information for which consent was not granted, the enforcement of privacy laws by courts could require a take-down of such information. Likewise, hosting and making available infringing content can expose platforms to liability and takedown, as evidenced by the Napster case and other cases where file-sharing platforms were enjoined by courts.

Those suing the providers of Generative AI today seek remedies including the takedown and retraining of AI models.[40]

There are known mechanisms for third-party intermediaries to address copyright violations, the most salient example being the Digital Millennium Copyright Act (DMCA) of the United States. Under this statute, third-party intermediaries can be shielded from liability if they create thorough mechanisms to investigate and act to address allegedly infringing content that they may be hosting. The DMCA exempts platforms from liability under a safe harbor provision through which they must follow so-called 'notice-and-takedown' rules. In short, 'notice-and-takedown' is a privately enforced system to address infringing content, through which anyone can file a complaint that is then arbitrated by the platform's designated agent. Platforms document this process in different ways, such as GitHub's DMCA Takedown Policy[41], HuggingFace's Content Policy[42] and GitLab's DMCA Policy[43].

Notice-and-takedown rules apply only to liability against copyright infringement in the United States. There are no clear legal standards with respect to harmful AI materials hosted by third-parties, and, absent clear mechanisms, these platforms may expose themselves to liability if they fail to take down content outside of the strict framework of US copyright law.

HuggingFace is a stellar example of community and platform led risk management and mitigation from AI harms. There are salient examples of HuggingFace removing harmful AI content and improving platform features to better identify and mitigate future harms from occurring. For example, in June 2023, Hugging Face hosted a text-to-image model called ERNIE ViLG, which was trained on the LAION-5B dataset, a large-scale multimodal dataset that contained numerous biases and harmful associations. Some users reported that the model generated stereotypical and offensive images based on the text prompts, such as depicting women as sexualized objects, or associating certain ethnicities with violence or poverty[44]. Hugging Face decided to take down the model from the platform, and issued a statement explaining their decision and apologizing for the harm caused. Similar examples arose with a language model called BLOOM[45] as well as a speech recognition model called Wav2Vec2[46].

Platforms require users to agree to terms of services and content policies, which represent an intent to shift the burden onto users for compliance with materials potentially triggering liability. Platforms can also create mechanisms for reporting and managing problematic AI materials, including DMCA mechanisms for copyright violations. However, these mechanisms typically require complaints from third parties. Platforms have a unique opportunity to devise technical means to implement M-type licensing. By doing so, platforms would give contributors the ability to effectively manage their contributions by modifying or revoking applicable licenses to their contributions. The authors believe that, if implemented soundly by third-party platforms, the mechanisms described in this article would further strengthen the case for responsible AI development. Pragmatically, platforms could also stand to gain by being able to point to additional means to empower creators to exercise content monitoring and control.

---

[40] *See footnote 32.*
[41] "DMCA Takedown Policy", GitHub Docs, accessed May 2024, https://docs.github.com/en/site-policy/content-removal-policies/dmca-takedown-policy.
[42] "Content Policy", Hugging face, August 30, 2023, https://huggingface.co/content-guidelines.
[43] Robin Schulman et al., "DMCA Policy", *The GitLab Handbook*, May 14, 2024, https://handbook.gitlab.com/handbook/legal/dmca/#dmca-digital-millennium-copyright-act-takedown-request-requirements.
[44] Leonardo Nicoletti, Dina Bass, "Humans Are Biased. Generative AI is Worse", *Bloomberg*, June 9, 2023, https://www.bloomberg.com/graphics/2023-generative-ai-bias/.
[45]
[46]

## 9. Conclusion

The authors fundamentally believe in the importance of open-science and open source software as a means to democratizing AI and furthering research to solve some of the fundamental issues we face as a society today. However, the pace of change in this space is nearly impossible to keep up with - whether it be in how technology can be leveraged for different purposes than originally imagined, evolution in testing and performance benchmarks (especially those related to bias and toxicity), changes to societal expectations around technology providers or other regulatory changes. The authors hope that M-licenses will be one of many tools that would ultimately help the community continuously strive for better, safer and fairer AI.

## *Appendix -* License text

**General Approach:**

Any license incorporating the M-type license language should be labeled as such (ex. M-MIT, M-BSD etc.) and contain the following header:

*This license grant is made under an M-Type license which is modifiable or revocable under certain circumstances as set out in the license text.*

In addition, license language will need to be added to the text of the applicable OSS license being used. The following language is license agnostic, but the best practice would be to use license-specific language that incorporates the definitions of the license itself. We set out license-specific language for some of the most common licenses below.

The grantor of this license may require that all licensees (i) cease using the contribution, (ii) update to a new version of the contribution, or (iii) accept certain use case restrictions regarding the contribution. Any notice by the grantor of this license of any modification or revocation of the license shall be immediately added to any notice file maintained by the licensee related to the contribution. The licensor's right to modify or revoke the license and its enforcement of that right can only be exercised in good faith under reasonably demonstrable assumptions in order to (i) address shortcomings of the contribution, such as a failure of applicable guardrails or model bias, (ii) prevent liability to the licensor, or (iii) prevent societal harms.

M-Apache 2.0 language

10. Modification or revocation of License. Subject to Section 11, the Licensor may require that all licensees (i) cease using their Contribution, (ii) update to a new version of the Contribution, or (iii) accept certain use case restrictions regarding the Contribution. Upon receipt of a notice by the Licensor of any modification or revocation of the License, you must (i) immediately add that NOTICE file wherever you have placed a NOTICE file pursuant to Section 4.d.
11. The Licensors' right to modify or revoke the License pursuant to Section 10 and its enforcement of that right can only be exercised in good faith under reasonably demonstrable assumptions in order to (i) address reasonably demonstrable shortcomings of the Work or Contribution, such as a failure of applicable guardrails or model bias, (ii) prevent liability to the Licensor, or (iii) prevent societal harms.

M-MIT language
<copyright holder> may require that all licensees (i) cease using the Software, (ii) update to a new version of the Software, or (iii) accept certain use case restrictions regarding the Software. Any notice by <copyright holder> of any modification or revocation of the license shall be immediately added to the permission notice required in all copies or substantial portions of the Software. <copyright holder>'s right to modify or revoke the license and its enforcement of that right can only be exercised in good faith under reasonably demonstrable assumptions in order to (i) address shortcomings of the Software, such as a failure of applicable guardrails or model bias, (ii) prevent liability to the copyright holder, or (iii) prevent societal harms.